\documentclass[osajnl,twocolumn,showpacs,superscriptaddress,10pt]{revtex4-1} 
\usepackage{amsmath,amssymb,graphicx}
\newcommand{\eryso}[0]{Er$^{3+}$:Y$_2$SiO$_5$ }
\newcommand{\tmyag}[0]{Tm$^{3+}$:YAG }

\newcommand{\yso}[0]{Y$_2$SiO$_5$ }

\newcommand{\ols}{\Omega_\mathrm{LS}}

\begin{document}

\title{Light-shift modulated photon-echo}

\author{Thierry Chaneli\`ere}
\affiliation{Laboratoire Aim\'e Cotton, CNRS, Universit\'e Paris-Sud and ENS Cachan, CNRS-UPR 3321, 91405 Orsay, France}
\author{Gabriel H\'etet}
\affiliation{Laboratoire Aim\'e Cotton, CNRS, Universit\'e Paris-Sud and ENS Cachan, CNRS-UPR 3321, 91405 Orsay, France}
\affiliation{Laboratoire Pierre Aigrain, Ecole Normale Sup\'erieure-PSL Research University, CNRS, Universit\'e Pierre et Marie Curie-Sorbonne Universit\'es, Universit\'e Paris Diderot-Sorbonne Paris Cit\'e, 24 rue Lhomond, 75231 Paris Cedex 05, France}

\begin{abstract}
We show that the AC-Stark shift (light-shift) is a powerful and versatile tool to control the emission of a photon-echo in the context of optical storage. As a proof-of-principle, we demonstrate that the photon-echo efficiency can be fully modulated by applying light-shift control pulses in an erbium doped solid. The control of the echo emission is attributed to the spatial gradient induced by the light-shift beam.
\end{abstract}

\ocis{(020.1670); (160.5690); (160.2900); (210.4680); (270.5565); (270.5585)}
\maketitle 

The photon-echo technique has been reconsidered recently with important applications in the context of quantum information storage and processing \cite{Bussieres}. The two-pulse or three-pulse schemes have inspired a variety of storage protocols 
\cite{tittel-photon}. The echo techniques have been implemented in different systems from atomic vapors to doped solids with remarkable performances in terms of efficiency \cite{Hosseini2011, Hedges2010}, bandwidth \cite{Saglamyurek2011}, multiplexing capacity \cite{bonarota11,sinclair14}. The two-pulse echo is indeed a stimulating source of inspiration to propose new protocols. In the echo sequence, the classical $\pi$-pulses must be associated with an extra control parameter to make the protocol suitable for quantum storage. This can be a rapidly switched electric field \cite{CRIB1, CRIB4, hetet08, Hyper}, a magnetic field \cite{wang1990modulation, Hetet:08,1367-2630-15-4-045015}, a modified phase-matching condition \cite{ROSE} or a frequency tunable active cavity \cite{Julsgaard, Wilson}.
These proposals cover different realities, atomic vapors and doped solids, both in the optical or the radio-frequency domain. The AC-Stark shift or light-shift can naturally complete this panoply for materials with a weak or zero sensitivity to the DC-Stark or Zeeman effects as pointed out early by Kraus {\it et al.} \cite{CRIB4}. As a counterpart of the DC-Stark shift which has already proven to be very efficient for quantum memories based on the so-called gradient echo protocol \cite{Hedges2010}, the light-shift is considered because of its versatility for the gradient design \cite{hetet_spinW} and its fast switching time \cite{ac_stark_gem}.  The imprinted phase pattern can also apply for optical memories based on ``electromagnetically induced transparency" \cite{PhysRevLett.103.033003}. This latter result demonstrates the deflection of the retrieved signal which is clearly a possibility offered by the light-shift gradient design. The shift induced by strong laser pulses on spin transitions (electronic and/or nuclear) is also used in spin-echo sequences showing the omnipresence of the phenomenon in different fields \cite{PhysRevA.42.1839,PhysRevA.78.013402, Berezovsky18042008}.

Within this general framework, we investigate experimentally the use of light-shift pulses in a standard two-pulse photon-echo sequence. We show that the emission can be fully modulated by applying light-shift pulses. Our demonstration is in that sense equivalent to the DC-Stark shift modulated spectroscopy pioneered by Meixner {\it et al.} \cite{PhysRevB.46.5912}. In our case, we attribute the modulation to the phase spatial pattern imprinted on the coherences by the light-shift beam. This phase gradient is a starting point to realize an  all-optical version of the gradient-echo memory \cite{ac_stark_gem, PhysRevLett.113.123602}. We first show that the emission of a two-pulse photon-echo (2PE) in \eryso can be controlled by applying a strong off-resonant pulse. The latter produces a light-shift during the free evolution of the coherences partially inhibiting the echo emission later on. We compensate this extra dephasing by a second light-shift pulse thus validating the method for controlling the emission of an optical memory.

\begin{figure}[h!]
\centering
\includegraphics[width=0.35\textwidth]{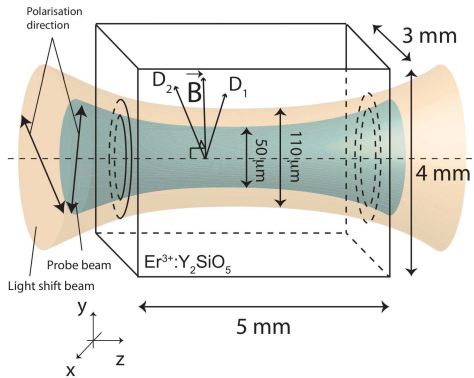}
\caption{\yso sample doped with 50 ppm of Er$^{3+}$. At 1.8 K and under a 2T magnetic field  in the plane ({\bf D$_1$}-{\bf D$_2$}), the coherence time is $\sim 130 \mu s$ ($^{4}$I$_{15/2}$ - $^{4}$I$_{13/2}$ transition for ``site 1'' ) \cite{dajczgewand2014optical,Bottger09}.}
\label{fig:setup}
\end{figure}

\begin{figure}[h!]
\centering
\includegraphics[width=0.4\textwidth]{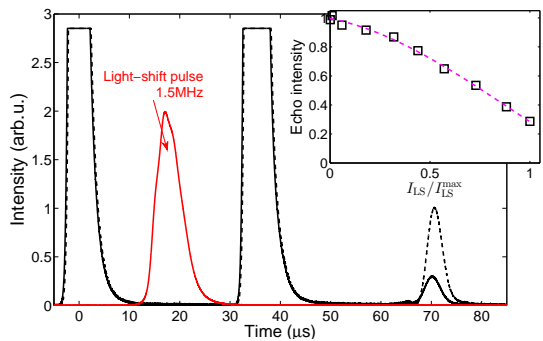}
\caption{Two-pulse photon echo sequence (dashed black line). We apply two strong pulses at $t=0$ and $t_{12}=35\mu s$ (clipped by the oscilloscope scale). We observe an echo at $2t_{12}=70\mu s$. The intensity is normalized so that the echo amplitude is $1$. When a light-shift pulse detuned by $1.5$MHz (in solid red) is applied at $t_{12}/2=17.5\mu s$, the echo intensity (solid black) is reduced from 100\% to 30\%. Inset: Reduction of the echo intensity as a function of the light-shift intensity $I_\mathrm{LS}$ (the dashed line is used to guide the eye). Measurement errors are a few percents given by shot-to-shot fluctuations due to the laser jitter (they roughly correspond to the markers size).}
\label{fig:time_seq_article}
\end{figure}

We choose \eryso as a test-bed for the light-shift modulated photon-echo experiment because it is recalcitrant to the efficient implementation of the gradient-echo scheme with DC-Stark shifts \cite{Lauritzen2011}. Our experimental setup has been extensively described previously in refs.\cite{Dajczgewand14, dajczgewand2014optical} (see Fig.\ref{fig:setup}). We implement a 2PE sequence in Fig.\ref{fig:time_seq_article} with a probe beam polarized along {\bf D$_1$} whose waist is 50 $\mu$m. Its  Rabi frequency is $2\pi \times 150$kHz measured by an optical nutation experiment \cite{allen1987ora}. There is no preparation of the medium by optical pumping. The sequence is repeated every 20ms so the atoms are all initially in the ground state. The 2PE is composed of two gaussian $1\mu s$ pulses (rms-duration) separated by  $t_{12}=35\mu s$. The echo is observed at $2t_{12}=70\mu s$ (Fig.\ref{fig:time_seq_article}). A second beam is used to produce off-resonant excitation (light-shift pulse) within the echo sequence. The latter has a waist of 110 $\mu$m and is polarized along {\bf D$_2$}. Its  Rabi frequency is $\ols^\mathrm{max}\simeq 2\pi \times$330kHz.  It is counterpropagating and overlapped with the probe beam.

In Fig.\ref{fig:time_seq_article}, we show that when a light-shift pulse whose rms-duration $\tau=3\mu s$ and detuned by $\Delta=2\pi \times 1.5$MHz is applied at $\frac{1}{2}t_{12}=17.5\mu s$, the echo is reduced  from 100\% to 30\%. In Fig.\ref{fig:time_seq_article} (inset), we also increase gradually the light-shift pulse intensity $I_\mathrm{LS}$ from $0$ to its maximum value $I_\mathrm{LS}^\mathrm{max}$ corresponding to $\ols^\mathrm{max}$.

The destruction of the echo by an extra pulse inserted in the time sequence is not obviously attributed to the light-shift induced on the coherence rephasing. Nevertheless it should be noted that the light-shift beam is sufficiently detuned to produce only an off-resonant excitation on the atoms driven by the probe.  The probe pulses have a duration of $1\mu s$ so their bandwidth is typically $150$kHz very comparable to the Rabi frequency of $2\pi \times 150$kHz. This is significantly lower than the light-shift beam detuning $1.5$MHz.

We now investigate the light-shift dependency as a function of the experimental parameters. If the light- shift pulse is described by its time-varying Rabi frequency $ \Omega(t)$, one expects the transition of the atoms under the probe to be shifted by $ \frac{\Omega^2(t)}{\Delta}$. The accumulated phase $\Phi_\mathrm{LS}$ is then
\begin{equation}
\Phi_\mathrm{LS}=\int_t \frac{\Omega^2(t)}{\Delta} \mathrm{d}t =\sqrt{2\pi} \frac{\ols^2}{\Delta}\tau
\label{phase_LS}
\end{equation}
given by the gaussian pulse parameters, $\ols$ its amplitude and $\tau$ its duration. We investigate this dependency (Eq.\ref{phase_LS}) by first varying $\Delta$. The time sequence is the same as before (see Fig.\ref{fig:time_seq_article}) expect that $\tau=2\mu s$ and $\Delta$ is varied from $2\pi \times 1$MHz to $2\pi \times 3$MHz.

\begin{figure}[h!]
\centering
\includegraphics[width=0.44\textwidth]{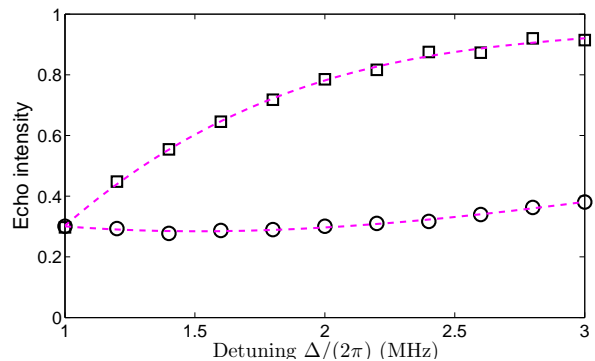}
\caption{Echo intensity when the light-shift pulse detuning $\Delta$ is varied (square symbols). The experiment is repeated by varying $\Delta$ but by keeping $\tau/\Delta$=$2\mu s/$MHz constant so $\tau$ goes from $2\mu s$ to  $6\mu s$ (circles). The dashed lines are used to guide the eye. Measurement errors are again a few percents.}
\label{fig:DeltaTau_article}
\end{figure}

When $\Delta$ is increased (Fig.\ref{fig:DeltaTau_article}), the effect of the light-shift pulse is reduced, thus qualitatively following the ${1}/{\Delta}$ dependency. A quantitative analysis is not directly possible because a perturbative treatment is inappropriate when strong pulses are used \cite{allen1987ora}. Alternatively, we propose to vary $\Delta$ but by keeping $\tau/\Delta$ constant to validate Eq.\eqref{phase_LS}. In this latter case, we observe that the effect of the light-shift is quasi-constant. A weak significant variation from 30\% to 38\% is still observable. It cannot be explained by Eq.\eqref{phase_LS}. Additional modeling would be required but the agreement with the expected $\tau/\Delta$ dependency is satisfying.

\begin{figure}[h!]
\centering
\includegraphics[width=0.4\textwidth]{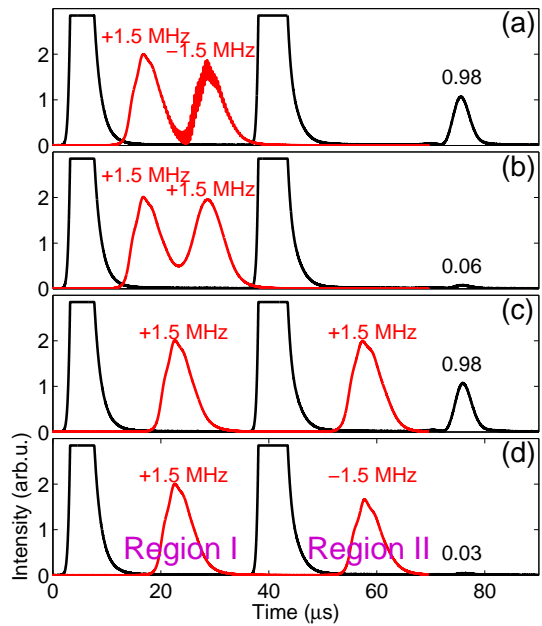}
\caption{Light-shift pulses compensation scheme. (a) If two pulses with opposite detunings are applied in region I, the echo intensity is 98\% of its initial value. (b) With the same detuning, the effect of the light-shift is cumulative and the echo is only 6\%. When applied in region I and II respectively, the pulses compensate each other with the same detuning (98\% echo  intensity in (c)) or add up with opposite detuning (3\% echo  intensity in (d)). The light-shift pulses have a $\pm1.5$MHz detuning and a $\tau=3\mu s$ duration (as in Fig.\ref{fig:time_seq_article}).}
\label{fig:serie_jeu_time_seq_article}
\end{figure}

To further explore this effect, we now propose to apply a first phase shift within the sequence and to compensate it by a second pulse. Looking at Eq.\eqref{phase_LS}, an intuitive compensation solution is to apply two successive pulses with opposite detunings during the free evolution between $t=0$ and $t_{12}$. A less obvious solution offered by the 2PE sequence is to apply one light-shift pulse between $t=0$ and $t_{12}$ (called region I) and a second one between $t_{12}$ and $2t_{12}$ (called region II) with the same detuning. 
To justify this compensation scheme, we can simply track down the accumulated phase $\phi (\omega, t)$ due to the inhomogeneous dephasing  \cite{allen1987ora}: {\bf (i) } In region I, after the first excitation, the coherence at the frequency $\omega$ freely evolves, accumulating $\phi (\omega, t)=\omega t$. Just before the second pulse at $t_{12}$, the phase is $\phi (\omega, t_{12}^-)=\omega t_{12}$. {\bf(ii)} If a light-shift pulse is  applied in region I, an extra term $\Phi_\mathrm{LS}^\mathrm{I}$ is added: $\phi (\omega, t_{12}^-)=\omega t_{12}+\Phi_\mathrm{LS}$. {\bf(iii)} The second pulse conjugates the coherence so the phase is now $\phi (\omega, t_{12}^+)=-\omega t_{12}-\Phi_\mathrm{LS}^\mathrm{I}$ right after the second pulse. {\bf(iv)} During the free evolution from $t_{12}$ to $t$  in region II, the accumulated phase is $\omega (t-t_{12})$ so the total phase is $\phi (\omega, t)=-\omega t_{12} -\Phi_\mathrm{LS}^\mathrm{I} +\omega (t-t_{12})=\omega (t-2t_{12})- \Phi_\mathrm{LS}^\mathrm{I}$. As expected, the retrieval time $2 t_{12}$ corresponds to the coherence rephasing. {\bf(v)} If a light-shift pulse is  applied in region II, an extra term $\Phi_\mathrm{LS}^\mathrm{II}$ is added. Thus the total inhomogeneous phase at the instant of retrieval is
\begin{equation}
\phi (\omega, t)=\omega (t-2t_{12}) -\Phi_\mathrm{LS}^\mathrm{I} +\Phi_\mathrm{LS}^\mathrm{II}.
\label{phase_inhom}\end{equation}
As a conclusion, similar pulses (same detuning) applied in region I and II compensate each other. Pulses with opposite detunings cancels each other only if they are both in region I or II exclusively. As we see in Fig.\ref{fig:serie_jeu_time_seq_article}, by properly choosing the sign of the detuning and the region of application, we can retrieve an echo with 98\% (Fig.(\ref{fig:serie_jeu_time_seq_article}.a) and (\ref{fig:serie_jeu_time_seq_article}.c)) of its initial reference intensity.

As a summary, the dependency as $\tau/\Delta$ illustrated in Fig.\ref{fig:DeltaTau_article} and the compensation scheme presented in Fig.\ref{fig:serie_jeu_time_seq_article} based on Eq.\eqref{phase_inhom} is a strong evidence that the echo is indeed controlled by the light-shift induced by the off-resonant pulses. This analysis justifies the main claim of our paper. Even if we have shown that the induced light-shift can be used to fully modulate the echo from 100\% to  3\%  (Fig.\ref{fig:serie_jeu_time_seq_article}.d), a quantitative link between Eq.\eqref{phase_LS} and the echo amplitude including gaussian propagation effects is not obvious. In a first approach, the echo amplitude is the total contribution of the excited atoms (at frequency $\omega$) at a given position $\mathbf{R}$ leading to an emission in the $\mathbf{k}$ direction proportional to  \cite{allen1987ora}:
\begin{equation}
\sum_{\omega,\vec{R}}\exp\left (\mathbf{k}.\mathbf{R}-\mathbf{k_\mathrm{in}}.\mathbf{R}+\phi (\omega, t)\right)
\label{echo_emission}\end{equation}
where $\mathbf{k_\mathrm{in}}$ is the probe wave-vector. The inhomogeneous phase $\phi (\omega, t)$  may also include a spatial dependency if the light-shifts depend on $\mathbf{R}$. Without light-shift, the echo peaks in the $\mathbf{k_\mathrm{in}}$ direction (phase-matching) at $t=2t_{12}$ (eq.\ref{phase_inhom}). It is important to note that a net global added phase to all the atoms excited by the 2PE does not change the echo amplitude but only its phase. The echo is reduced only if the phase varies through the inhomogeneous profile (depends on $\omega$, spectral dependency) or is not constant through the sample ($\mathbf{R}$, spatial dependency). A spectral dependency may prevent the coherence rephasing at $t=2t_{12}$. A spatial dependency modifies the phase-matching condition when the echo is emitted. We discuss these two possible effects before concluding.

A spectral dependency induced by the light-shift has a negligible influence in our case because the detuning is significantly larger than the excited bandwidth as previously mentioned. It should be noted that the first order effect of a spectral dependency described by
\begin{equation}
\Phi_\mathrm{LS}(\omega)=\sqrt{2\pi} \frac{\ols^2}{\Delta+\omega}\tau \simeq \sqrt{2\pi} \frac{\ols^2}{\Delta}\tau  - \sqrt{2\pi} \frac{\ols^2}{\Delta^2}\tau \omega
\label{phase_LS_spectral}
\end{equation}
is to modify the retrieval time from $2t_{12}$ to  $ 2t_{12}\pm\sqrt{2\pi} \tau{\ols^2}/{\Delta^2}$ (see Eq.\ref{phase_inhom} depending if the pulse is in region I or II). The pulse should be delayed to the first order. This is not what we observe so the spectral dependency is certainly negligible.

The spatial dependency of the induced light-shift $\Phi_\mathrm{LS}(z,r)$ can now be discussed. It may be both longitudinal (along $z$) and transverse (along $r$, in the $(x,y)$ plane). We choose to overlap the probe and light-shift beams, as a consequence the longitudinal and transverse gradients are decoupled (rotational symmetry). It simplifies our analysis at this level. Nevertheless, we have also verified experimentally that angled beams produces a significant reduction of the photon-echo. It should be kept in mind when more practical conditions are considered. In our case, the $z$-gradient may be due to the absorption of the light-shift beam $\ols(z)$ along the propagation direction $z$. The $r$-gradient appears because the light-shift beam as a finite size (110 $\mu$m) with respect to the probe (50 $\mu$m). The longitudinal dependency can be evaluated. At $z=0$, we have $\ols(0)\simeq 2\pi \times$330kHz and then $\Phi_\mathrm{LS}(0)\simeq 1.1 \pi$. The gaussian divergence of the light-shift beam is negligible because its Rayleigh length is ten times longer than the crystal. On the contrary, its complete absorption would produce a phase gradient from  $1.1 \pi$ at $z=0$ to $0\pi$ at the output of the sample. In other words, the input and the output slice emissions would be out of phase thus explaining the echo reduction (phase mismatch). This appealing explanation is unfortunately extremely unlikely.  The medium is absorbing for the light-shift beam which is polarized along  {\bf D$_2$} (optical depth $\sim$ 3.5). Nevertheless, the light-shift pulses area is large ($\sim 7 \pi$) so they are not absorbed as small-area pulses would be \cite{allen1987ora}. They may exhibit solitonic propagation or self-induced-transparency, in any case their amplitude will not go to zero. We have performed 1D-Bloch-Maxwell numerical simulations modeling the echo emission and the light-shift pulse propagation along $z$ thus taking into account the spectral and the longitudinal spatial dependency. We could not simulate the echo reduction but only the modified echo retrieval time by $\sqrt{2\pi}\tau {\ols^2}/{\Delta^2}$ which is not the dominant experimental observation anyway. The simulated light-shift  pulse propagation qualitatively corresponds to the experimental outgoing shape predicting a negligible $\sim$3\% absorption for a $7 \pi$ area pulse. The transverse spatial dependency is a very likely explanation. A first order estimation of the transverse shift induced by the probe (50 $\mu$m) and light shift beam (110 $\mu$m) spatial mode mismatch only predicts a 10\% echo modulation. 3D-Bloch-Maxwell numerical simulations would be necessary to account for the three-dimensional propagation of the echo.

To conclude, we show that the photon echo emission can be fully modulated by applying off-resonant pulses within the time sequence. The echo is modified by the transverse spatial phase gradient imprinted by the light-shift beam on the atomic coherence.. It is a powerful and versatile tool to manipulate the emission of quantum memories.  We obtain very comparable results with a \tmyag sample showing the entirety of the phenomenon. Our demonstration opens up new perspectives for materials with a low Stark or Zeeman sensitivity.

We have received funding from the Marie Curie Actions of the European Union's 7th Framework Programme under REA no. 287252, from the national grants ANR-12-BS08-0015-02 (RAMACO) and  ANR-13-PDOC-0024-01 (retour post-doctorants SMEQUI).
\bibliography{LS2PE_bib_resub}

\begin{thebibliography}{30}%
\makeatletter
\providecommand \@ifxundefined [1]{%
 \@ifx{#1\undefined}
}%
\providecommand \@ifnum [1]{%
 \ifnum #1\expandafter \@firstoftwo
 \else \expandafter \@secondoftwo
 \fi
}%
\providecommand \@ifx [1]{%
 \ifx #1\expandafter \@firstoftwo
 \else \expandafter \@secondoftwo
 \fi
}%
\providecommand \natexlab [1]{#1}%
\providecommand \enquote  [1]{``#1''}%
\providecommand \bibnamefont  [1]{#1}%
\providecommand \bibfnamefont [1]{#1}%
\providecommand \citenamefont [1]{#1}%
\providecommand \href@noop [0]{\@secondoftwo}%
\providecommand \href [0]{\begingroup \@sanitize@url \@href}%
\providecommand \@href[1]{\@@startlink{#1}\@@href}%
\providecommand \@@href[1]{\endgroup#1\@@endlink}%
\providecommand \@sanitize@url [0]{\catcode `\\12\catcode `\$12\catcode
  `\&12\catcode `\#12\catcode `\^12\catcode `\_12\catcode `\%12\relax}%
\providecommand \@@startlink[1]{}%
\providecommand \@@endlink[0]{}%
\providecommand \url  [0]{\begingroup\@sanitize@url \@url }%
\providecommand \@url [1]{\endgroup\@href {#1}{\urlprefix }}%
\providecommand \urlprefix  [0]{URL }%
\providecommand \Eprint [0]{\href }%
\providecommand \doibase [0]{http://dx.doi.org/}%
\providecommand \selectlanguage [0]{\@gobble}%
\providecommand \bibinfo  [0]{\@secondoftwo}%
\providecommand \bibfield  [0]{\@secondoftwo}%
\providecommand \translation [1]{[#1]}%
\providecommand \BibitemOpen [0]{}%
\providecommand \bibitemStop [0]{}%
\providecommand \bibitemNoStop [0]{.\EOS\space}%
\providecommand \EOS [0]{\spacefactor3000\relax}%
\providecommand \BibitemShut  [1]{\csname bibitem#1\endcsname}%
\let\auto@bib@innerbib\@empty
\bibitem [{\citenamefont {Bussi\`eres}\ \emph {et~al.}(2013)\citenamefont
  {Bussi\`eres}, \citenamefont {Sangouard}, \citenamefont {Afzelius},
  \citenamefont {de~Riedmatten}, \citenamefont {Simon},\ and\ \citenamefont
  {Tittel}}]{Bussieres}%
  \BibitemOpen
  \bibfield  {author} {\bibinfo {author} {\bibfnamefont {F.}~\bibnamefont
  {Bussi\`eres}}, \bibinfo {author} {\bibfnamefont {N.}~\bibnamefont
  {Sangouard}}, \bibinfo {author} {\bibfnamefont {M.}~\bibnamefont {Afzelius}},
  \bibinfo {author} {\bibfnamefont {H.}~\bibnamefont {de~Riedmatten}}, \bibinfo
  {author} {\bibfnamefont {C.}~\bibnamefont {Simon}}, \ and\ \bibinfo {author}
  {\bibfnamefont {W.}~\bibnamefont {Tittel}},\ }\href {\doibase
  10.1080/09500340.2013.856482} {\bibfield  {journal} {\bibinfo  {journal} {J.
  Mod. Opt.}\ }\textbf {\bibinfo {volume} {60}},\ \bibinfo {pages} {1519}
  (\bibinfo {year} {2013})}\BibitemShut {NoStop}%
\bibitem [{\citenamefont {Tittel}\ \emph {et~al.}(2010)\citenamefont {Tittel},
  \citenamefont {Afzelius}, \citenamefont {Cone}, \citenamefont
  {Chaneli{\`e}re}, \citenamefont {Kr{\"o}ll}, \citenamefont {Moiseev},\ and\
  \citenamefont {Sellars}}]{tittel-photon}%
  \BibitemOpen
  \bibfield  {author} {\bibinfo {author} {\bibfnamefont {W.}~\bibnamefont
  {Tittel}}, \bibinfo {author} {\bibfnamefont {M.}~\bibnamefont {Afzelius}},
  \bibinfo {author} {\bibfnamefont {R.}~\bibnamefont {Cone}}, \bibinfo {author}
  {\bibfnamefont {T.}~\bibnamefont {Chaneli{\`e}re}}, \bibinfo {author}
  {\bibfnamefont {S.}~\bibnamefont {Kr{\"o}ll}}, \bibinfo {author}
  {\bibfnamefont {S.}~\bibnamefont {Moiseev}}, \ and\ \bibinfo {author}
  {\bibfnamefont {M.}~\bibnamefont {Sellars}},\ }\href@noop {} {\bibfield
  {journal} {\bibinfo  {journal} {Laser Photon. Rev.}\ }\textbf {\bibinfo
  {volume} {4}},\ \bibinfo {pages} {244} (\bibinfo {year} {2010})}\BibitemShut
  {NoStop}%
\bibitem [{\citenamefont {Hosseini}\ \emph {et~al.}(2011)\citenamefont
  {Hosseini}, \citenamefont {Sparkes}, \citenamefont {Campbell}, \citenamefont
  {Lam},\ and\ \citenamefont {Buchler}}]{Hosseini2011}%
  \BibitemOpen
  \bibfield  {author} {\bibinfo {author} {\bibfnamefont {M.}~\bibnamefont
  {Hosseini}}, \bibinfo {author} {\bibfnamefont {B.~M.}\ \bibnamefont
  {Sparkes}}, \bibinfo {author} {\bibfnamefont {G.}~\bibnamefont {Campbell}},
  \bibinfo {author} {\bibfnamefont {P.~K.}\ \bibnamefont {Lam}}, \ and\
  \bibinfo {author} {\bibfnamefont {B.~C.}\ \bibnamefont {Buchler}},\ }\href
  {\doibase 10.1038/ncomms1175} {\bibfield  {journal} {\bibinfo  {journal}
  {Nat. Commun.}\ }\textbf {\bibinfo {volume} {2}},\ \bibinfo {pages} {174}
  (\bibinfo {year} {2011})}\BibitemShut {NoStop}%
\bibitem [{\citenamefont {Hedges}\ \emph {et~al.}(2010)\citenamefont {Hedges},
  \citenamefont {Longdell}, \citenamefont {Li},\ and\ \citenamefont
  {Sellars}}]{Hedges2010}%
  \BibitemOpen
  \bibfield  {author} {\bibinfo {author} {\bibfnamefont {M.~P.}\ \bibnamefont
  {Hedges}}, \bibinfo {author} {\bibfnamefont {J.~J.}\ \bibnamefont
  {Longdell}}, \bibinfo {author} {\bibfnamefont {Y.}~\bibnamefont {Li}}, \ and\
  \bibinfo {author} {\bibfnamefont {M.~J.}\ \bibnamefont {Sellars}},\ }\href
  {\doibase 10.1038/nature09081} {\bibfield  {journal} {\bibinfo  {journal}
  {Nature}\ }\textbf {\bibinfo {volume} {465}},\ \bibinfo {pages} {1052}
  (\bibinfo {year} {2010})}\BibitemShut {NoStop}%
\bibitem [{\citenamefont {Saglamyurek}\ \emph {et~al.}(2011)\citenamefont
  {Saglamyurek}, \citenamefont {Sinclair}, \citenamefont {Jin}, \citenamefont
  {Slater}, \citenamefont {Oblak}, \citenamefont {Bussieres}, \citenamefont
  {George}, \citenamefont {Ricken}, \citenamefont {Sohler},\ and\ \citenamefont
  {Tittel}}]{Saglamyurek2011}%
  \BibitemOpen
  \bibfield  {author} {\bibinfo {author} {\bibfnamefont {E.}~\bibnamefont
  {Saglamyurek}}, \bibinfo {author} {\bibfnamefont {N.}~\bibnamefont
  {Sinclair}}, \bibinfo {author} {\bibfnamefont {J.}~\bibnamefont {Jin}},
  \bibinfo {author} {\bibfnamefont {J.~A.}\ \bibnamefont {Slater}}, \bibinfo
  {author} {\bibfnamefont {D.}~\bibnamefont {Oblak}}, \bibinfo {author}
  {\bibfnamefont {F.}~\bibnamefont {Bussieres}}, \bibinfo {author}
  {\bibfnamefont {M.}~\bibnamefont {George}}, \bibinfo {author} {\bibfnamefont
  {R.}~\bibnamefont {Ricken}}, \bibinfo {author} {\bibfnamefont
  {W.}~\bibnamefont {Sohler}}, \ and\ \bibinfo {author} {\bibfnamefont
  {W.}~\bibnamefont {Tittel}},\ }\href {\doibase 10.1038/nature09719}
  {\bibfield  {journal} {\bibinfo  {journal} {Nature}\ }\textbf {\bibinfo
  {volume} {469}},\ \bibinfo {pages} {512} (\bibinfo {year}
  {2011})}\BibitemShut {NoStop}%
\bibitem [{\citenamefont {Bonarota}\ \emph {et~al.}(2011)\citenamefont
  {Bonarota}, \citenamefont {Gou\"et},\ and\ \citenamefont
  {Chaneli\`ere}}]{bonarota11}%
  \BibitemOpen
  \bibfield  {author} {\bibinfo {author} {\bibfnamefont {M.}~\bibnamefont
  {Bonarota}}, \bibinfo {author} {\bibfnamefont {J.-L.~L.}\ \bibnamefont
  {Gou\"et}}, \ and\ \bibinfo {author} {\bibfnamefont {T.}~\bibnamefont
  {Chaneli\`ere}},\ }\href {http://stacks.iop.org/1367-2630/13/i=1/a=013013}
  {\bibfield  {journal} {\bibinfo  {journal} {New J. Phys.}\ }\textbf {\bibinfo
  {volume} {13}},\ \bibinfo {pages} {013013} (\bibinfo {year}
  {2011})}\BibitemShut {NoStop}%
\bibitem [{\citenamefont {Sinclair}\ \emph {et~al.}(2014)\citenamefont
  {Sinclair}, \citenamefont {Saglamyurek}, \citenamefont {Mallahzadeh},
  \citenamefont {Slater}, \citenamefont {George}, \citenamefont {Ricken},
  \citenamefont {Hedges}, \citenamefont {Oblak}, \citenamefont {Simon},
  \citenamefont {Sohler},\ and\ \citenamefont {Tittel}}]{sinclair14}%
  \BibitemOpen
  \bibfield  {author} {\bibinfo {author} {\bibfnamefont {N.}~\bibnamefont
  {Sinclair}}, \bibinfo {author} {\bibfnamefont {E.}~\bibnamefont
  {Saglamyurek}}, \bibinfo {author} {\bibfnamefont {H.}~\bibnamefont
  {Mallahzadeh}}, \bibinfo {author} {\bibfnamefont {J.~A.}\ \bibnamefont
  {Slater}}, \bibinfo {author} {\bibfnamefont {M.}~\bibnamefont {George}},
  \bibinfo {author} {\bibfnamefont {R.}~\bibnamefont {Ricken}}, \bibinfo
  {author} {\bibfnamefont {M.~P.}\ \bibnamefont {Hedges}}, \bibinfo {author}
  {\bibfnamefont {D.}~\bibnamefont {Oblak}}, \bibinfo {author} {\bibfnamefont
  {C.}~\bibnamefont {Simon}}, \bibinfo {author} {\bibfnamefont
  {W.}~\bibnamefont {Sohler}}, \ and\ \bibinfo {author} {\bibfnamefont
  {W.}~\bibnamefont {Tittel}},\ }\href {\doibase
  10.1103/PhysRevLett.113.053603} {\bibfield  {journal} {\bibinfo  {journal}
  {Phys. Rev. Lett.}\ }\textbf {\bibinfo {volume} {113}},\ \bibinfo {pages}
  {053603} (\bibinfo {year} {2014})}\BibitemShut {NoStop}%
\bibitem [{\citenamefont {Nilsson}\ and\ \citenamefont
  {Kr\"oll}(2005)}]{CRIB1}%
  \BibitemOpen
  \bibfield  {author} {\bibinfo {author} {\bibfnamefont {M.}~\bibnamefont
  {Nilsson}}\ and\ \bibinfo {author} {\bibfnamefont {S.}~\bibnamefont
  {Kr\"oll}},\ }\href {\doibase 10.1016/j.optcom.2004.11.077} {\bibfield
  {journal} {\bibinfo  {journal} {Optics Comm.}\ }\textbf {\bibinfo {volume}
  {247}},\ \bibinfo {pages} {393} (\bibinfo {year} {2005})}\BibitemShut
  {NoStop}%
\bibitem [{\citenamefont {Kraus}\ \emph {et~al.}(2006)\citenamefont {Kraus},
  \citenamefont {Tittel}, \citenamefont {Gisin}, \citenamefont {Nilsson},
  \citenamefont {Kr\"oll},\ and\ \citenamefont {Cirac}}]{CRIB4}%
  \BibitemOpen
  \bibfield  {author} {\bibinfo {author} {\bibfnamefont {B.}~\bibnamefont
  {Kraus}}, \bibinfo {author} {\bibfnamefont {W.}~\bibnamefont {Tittel}},
  \bibinfo {author} {\bibfnamefont {N.}~\bibnamefont {Gisin}}, \bibinfo
  {author} {\bibfnamefont {M.}~\bibnamefont {Nilsson}}, \bibinfo {author}
  {\bibfnamefont {S.}~\bibnamefont {Kr\"oll}}, \ and\ \bibinfo {author}
  {\bibfnamefont {J.~I.}\ \bibnamefont {Cirac}},\ }\href {\doibase
  10.1103/PhysRevA.73.020302} {\bibfield  {journal} {\bibinfo  {journal} {Phys.
  Rev. A}\ }\textbf {\bibinfo {volume} {73}},\ \bibinfo {pages} {020302}
  (\bibinfo {year} {2006})}\BibitemShut {NoStop}%
\bibitem [{\citenamefont {H\'etet}\ \emph {et~al.}(2008)\citenamefont
  {H\'etet}, \citenamefont {Longdell}, \citenamefont {Alexander}, \citenamefont
  {Lam},\ and\ \citenamefont {Sellars}}]{hetet08}%
  \BibitemOpen
  \bibfield  {author} {\bibinfo {author} {\bibfnamefont {G.}~\bibnamefont
  {H\'etet}}, \bibinfo {author} {\bibfnamefont {J.}~\bibnamefont {Longdell}},
  \bibinfo {author} {\bibfnamefont {A.}~\bibnamefont {Alexander}}, \bibinfo
  {author} {\bibfnamefont {P.}~\bibnamefont {Lam}}, \ and\ \bibinfo {author}
  {\bibfnamefont {M.}~\bibnamefont {Sellars}},\ }\href {\doibase
  10.1103/PhysRevLett.100.023601} {\bibfield  {journal} {\bibinfo  {journal}
  {Phys. Rev. Lett.}\ }\textbf {\bibinfo {volume} {100}},\ \bibinfo {pages}
  {023601} (\bibinfo {year} {2008})}\BibitemShut {NoStop}%
\bibitem [{\citenamefont {McAuslan}\ \emph {et~al.}(2011)\citenamefont
  {McAuslan}, \citenamefont {Ledingham}, \citenamefont {Naylor}, \citenamefont
  {Beavan}, \citenamefont {Hedges}, \citenamefont {Sellars},\ and\
  \citenamefont {Longdell}}]{Hyper}%
  \BibitemOpen
  \bibfield  {author} {\bibinfo {author} {\bibfnamefont {D.~L.}\ \bibnamefont
  {McAuslan}}, \bibinfo {author} {\bibfnamefont {P.~M.}\ \bibnamefont
  {Ledingham}}, \bibinfo {author} {\bibfnamefont {W.~R.}\ \bibnamefont
  {Naylor}}, \bibinfo {author} {\bibfnamefont {S.~E.}\ \bibnamefont {Beavan}},
  \bibinfo {author} {\bibfnamefont {M.~P.}\ \bibnamefont {Hedges}}, \bibinfo
  {author} {\bibfnamefont {M.~J.}\ \bibnamefont {Sellars}}, \ and\ \bibinfo
  {author} {\bibfnamefont {J.~J.}\ \bibnamefont {Longdell}},\ }\href {\doibase
  10.1103/PhysRevA.84.022309} {\bibfield  {journal} {\bibinfo  {journal} {Phys.
  Rev. A}\ }\textbf {\bibinfo {volume} {84}},\ \bibinfo {pages} {022309}
  (\bibinfo {year} {2011})}\BibitemShut {NoStop}%
\bibitem [{\citenamefont {Wang}\ \emph {et~al.}(1990)\citenamefont {Wang},
  \citenamefont {Boye}, \citenamefont {Rives},\ and\ \citenamefont
  {Meltzer}}]{wang1990modulation}%
  \BibitemOpen
  \bibfield  {author} {\bibinfo {author} {\bibfnamefont {Y.}~\bibnamefont
  {Wang}}, \bibinfo {author} {\bibfnamefont {D.}~\bibnamefont {Boye}}, \bibinfo
  {author} {\bibfnamefont {J.}~\bibnamefont {Rives}}, \ and\ \bibinfo {author}
  {\bibfnamefont {R.}~\bibnamefont {Meltzer}},\ }\href@noop {} {\bibfield
  {journal} {\bibinfo  {journal} {J. Lumin.}\ }\textbf {\bibinfo {volume}
  {45}},\ \bibinfo {pages} {437} (\bibinfo {year} {1990})}\BibitemShut
  {NoStop}%
\bibitem [{\citenamefont {H\'{e}tet}\ \emph {et~al.}(2008)\citenamefont
  {H\'{e}tet}, \citenamefont {Hosseini}, \citenamefont {Sparkes}, \citenamefont
  {Oblak}, \citenamefont {Lam},\ and\ \citenamefont {Buchler}}]{Hetet:08}%
  \BibitemOpen
  \bibfield  {author} {\bibinfo {author} {\bibfnamefont {G.}~\bibnamefont
  {H\'{e}tet}}, \bibinfo {author} {\bibfnamefont {M.}~\bibnamefont {Hosseini}},
  \bibinfo {author} {\bibfnamefont {B.~M.}\ \bibnamefont {Sparkes}}, \bibinfo
  {author} {\bibfnamefont {D.}~\bibnamefont {Oblak}}, \bibinfo {author}
  {\bibfnamefont {P.~K.}\ \bibnamefont {Lam}}, \ and\ \bibinfo {author}
  {\bibfnamefont {B.~C.}\ \bibnamefont {Buchler}},\ }\href {\doibase
  10.1364/OL.33.002323} {\bibfield  {journal} {\bibinfo  {journal} {Opt.
  Lett.}\ }\textbf {\bibinfo {volume} {33}},\ \bibinfo {pages} {2323} (\bibinfo
  {year} {2008})}\BibitemShut {NoStop}%
\bibitem [{\citenamefont {H\'etet}\ \emph {et~al.}(2013)\citenamefont
  {H\'etet}, \citenamefont {Wilkowski},\ and\ \citenamefont
  {Chaneli\`ere}}]{1367-2630-15-4-045015}%
  \BibitemOpen
  \bibfield  {author} {\bibinfo {author} {\bibfnamefont {G.}~\bibnamefont
  {H\'etet}}, \bibinfo {author} {\bibfnamefont {D.}~\bibnamefont {Wilkowski}},
  \ and\ \bibinfo {author} {\bibfnamefont {T.}~\bibnamefont {Chaneli\`ere}},\
  }\href {http://stacks.iop.org/1367-2630/15/i=4/a=045015} {\bibfield
  {journal} {\bibinfo  {journal} {New J. Phys.}\ }\textbf {\bibinfo {volume}
  {15}},\ \bibinfo {pages} {045015} (\bibinfo {year} {2013})}\BibitemShut
  {NoStop}%
\bibitem [{\citenamefont {Damon}\ \emph {et~al.}(2011)\citenamefont {Damon},
  \citenamefont {Bonarota}, \citenamefont {Louchet-Chauvet}, \citenamefont
  {Chaneli\`ere},\ and\ \citenamefont {Le~Gou\"et}}]{ROSE}%
  \BibitemOpen
  \bibfield  {author} {\bibinfo {author} {\bibfnamefont {V.}~\bibnamefont
  {Damon}}, \bibinfo {author} {\bibfnamefont {M.}~\bibnamefont {Bonarota}},
  \bibinfo {author} {\bibfnamefont {A.}~\bibnamefont {Louchet-Chauvet}},
  \bibinfo {author} {\bibfnamefont {T.}~\bibnamefont {Chaneli\`ere}}, \ and\
  \bibinfo {author} {\bibfnamefont {J.-L.}\ \bibnamefont {Le~Gou\"et}},\
  }\href@noop {} {\bibfield  {journal} {\bibinfo  {journal} {New J. Phys.}\
  }\textbf {\bibinfo {volume} {13}},\ \bibinfo {pages} {093031} (\bibinfo
  {year} {2011})}\BibitemShut {NoStop}%
\bibitem [{\citenamefont {Julsgaard}\ \emph {et~al.}(2013)\citenamefont
  {Julsgaard}, \citenamefont {Grezes}, \citenamefont {Bertet},\ and\
  \citenamefont {M\o{}lmer}}]{Julsgaard}%
  \BibitemOpen
  \bibfield  {author} {\bibinfo {author} {\bibfnamefont {B.}~\bibnamefont
  {Julsgaard}}, \bibinfo {author} {\bibfnamefont {C.}~\bibnamefont {Grezes}},
  \bibinfo {author} {\bibfnamefont {P.}~\bibnamefont {Bertet}}, \ and\ \bibinfo
  {author} {\bibfnamefont {K.}~\bibnamefont {M\o{}lmer}},\ }\href {\doibase
  10.1103/PhysRevLett.110.250503} {\bibfield  {journal} {\bibinfo  {journal}
  {Phys. Rev. Lett.}\ }\textbf {\bibinfo {volume} {110}},\ \bibinfo {pages}
  {250503} (\bibinfo {year} {2013})}\BibitemShut {NoStop}%
\bibitem [{\citenamefont {Afzelius}\ \emph {et~al.}(2013)\citenamefont
  {Afzelius}, \citenamefont {Sangouard}, \citenamefont {Johansson},
  \citenamefont {Staudt},\ and\ \citenamefont {Wilson}}]{Wilson}%
  \BibitemOpen
  \bibfield  {author} {\bibinfo {author} {\bibfnamefont {M.}~\bibnamefont
  {Afzelius}}, \bibinfo {author} {\bibfnamefont {N.}~\bibnamefont {Sangouard}},
  \bibinfo {author} {\bibfnamefont {G.}~\bibnamefont {Johansson}}, \bibinfo
  {author} {\bibfnamefont {M.~U.}\ \bibnamefont {Staudt}}, \ and\ \bibinfo
  {author} {\bibfnamefont {C.~M.}\ \bibnamefont {Wilson}},\ }\href
  {http://stacks.iop.org/1367-2630/15/i=6/a=065008} {\bibfield  {journal}
  {\bibinfo  {journal} {New J. Phys.}\ }\textbf {\bibinfo {volume} {15}},\
  \bibinfo {pages} {065008} (\bibinfo {year} {2013})}\BibitemShut {NoStop}%
\bibitem [{\citenamefont {H\'etet}\ and\ \citenamefont
  {Gu\'ery-Odelin}()}]{hetet_spinW}%
  \BibitemOpen
  \bibfield  {author} {\bibinfo {author} {\bibfnamefont {G.}~\bibnamefont
  {H\'etet}}\ and\ \bibinfo {author} {\bibfnamefont {D.}~\bibnamefont
  {Gu\'ery-Odelin}},\ }\href@noop {} {\ }\Eprint
  {http://arxiv.org/abs/quant-ph/1501.01194} {arXiv:quant-ph/1501.01194}
  \BibitemShut {NoStop}%
\bibitem [{\citenamefont {Sparkes}\ \emph {et~al.}(2010)\citenamefont
  {Sparkes}, \citenamefont {Hosseini}, \citenamefont {H\'etet}, \citenamefont
  {Lam},\ and\ \citenamefont {Buchler}}]{ac_stark_gem}%
  \BibitemOpen
  \bibfield  {author} {\bibinfo {author} {\bibfnamefont {B.~M.}\ \bibnamefont
  {Sparkes}}, \bibinfo {author} {\bibfnamefont {M.}~\bibnamefont {Hosseini}},
  \bibinfo {author} {\bibfnamefont {G.}~\bibnamefont {H\'etet}}, \bibinfo
  {author} {\bibfnamefont {P.~K.}\ \bibnamefont {Lam}}, \ and\ \bibinfo
  {author} {\bibfnamefont {B.~C.}\ \bibnamefont {Buchler}},\ }\href {\doibase
  10.1103/PhysRevA.82.043847} {\bibfield  {journal} {\bibinfo  {journal} {Phys.
  Rev. A}\ }\textbf {\bibinfo {volume} {82}},\ \bibinfo {pages} {043847}
  (\bibinfo {year} {2010})}\BibitemShut {NoStop}%
\bibitem [{\citenamefont {Schnorrberger}\ \emph {et~al.}(2009)\citenamefont
  {Schnorrberger}, \citenamefont {Thompson}, \citenamefont {Trotzky},
  \citenamefont {Pugatch}, \citenamefont {Davidson}, \citenamefont {Kuhr},\
  and\ \citenamefont {Bloch}}]{PhysRevLett.103.033003}%
  \BibitemOpen
  \bibfield  {author} {\bibinfo {author} {\bibfnamefont {U.}~\bibnamefont
  {Schnorrberger}}, \bibinfo {author} {\bibfnamefont {J.~D.}\ \bibnamefont
  {Thompson}}, \bibinfo {author} {\bibfnamefont {S.}~\bibnamefont {Trotzky}},
  \bibinfo {author} {\bibfnamefont {R.}~\bibnamefont {Pugatch}}, \bibinfo
  {author} {\bibfnamefont {N.}~\bibnamefont {Davidson}}, \bibinfo {author}
  {\bibfnamefont {S.}~\bibnamefont {Kuhr}}, \ and\ \bibinfo {author}
  {\bibfnamefont {I.}~\bibnamefont {Bloch}},\ }\href {\doibase
  10.1103/PhysRevLett.103.033003} {\bibfield  {journal} {\bibinfo  {journal}
  {Phys. Rev. Lett.}\ }\textbf {\bibinfo {volume} {103}},\ \bibinfo {pages}
  {033003} (\bibinfo {year} {2009})}\BibitemShut {NoStop}%
\bibitem [{\citenamefont {Rosatzin}\ \emph {et~al.}(1990)\citenamefont
  {Rosatzin}, \citenamefont {Suter},\ and\ \citenamefont
  {Mlynek}}]{PhysRevA.42.1839}%
  \BibitemOpen
  \bibfield  {author} {\bibinfo {author} {\bibfnamefont {M.}~\bibnamefont
  {Rosatzin}}, \bibinfo {author} {\bibfnamefont {D.}~\bibnamefont {Suter}}, \
  and\ \bibinfo {author} {\bibfnamefont {J.}~\bibnamefont {Mlynek}},\ }\href
  {\doibase 10.1103/PhysRevA.42.1839} {\bibfield  {journal} {\bibinfo
  {journal} {Phys. Rev. A}\ }\textbf {\bibinfo {volume} {42}},\ \bibinfo
  {pages} {1839} (\bibinfo {year} {1990})}\BibitemShut {NoStop}%
\bibitem [{\citenamefont {Moriyasu}\ \emph {et~al.}(2008)\citenamefont
  {Moriyasu}, \citenamefont {Koyama}, \citenamefont {Fukuda},\ and\
  \citenamefont {Kohmoto}}]{PhysRevA.78.013402}%
  \BibitemOpen
  \bibfield  {author} {\bibinfo {author} {\bibfnamefont {T.}~\bibnamefont
  {Moriyasu}}, \bibinfo {author} {\bibfnamefont {Y.}~\bibnamefont {Koyama}},
  \bibinfo {author} {\bibfnamefont {Y.}~\bibnamefont {Fukuda}}, \ and\ \bibinfo
  {author} {\bibfnamefont {T.}~\bibnamefont {Kohmoto}},\ }\href {\doibase
  10.1103/PhysRevA.78.013402} {\bibfield  {journal} {\bibinfo  {journal} {Phys.
  Rev. A}\ }\textbf {\bibinfo {volume} {78}},\ \bibinfo {pages} {013402}
  (\bibinfo {year} {2008})}\BibitemShut {NoStop}%
\bibitem [{\citenamefont {Berezovsky}\ \emph {et~al.}(2008)\citenamefont
  {Berezovsky}, \citenamefont {Mikkelsen}, \citenamefont {Stoltz},
  \citenamefont {Coldren},\ and\ \citenamefont
  {Awschalom}}]{Berezovsky18042008}%
  \BibitemOpen
  \bibfield  {author} {\bibinfo {author} {\bibfnamefont {J.}~\bibnamefont
  {Berezovsky}}, \bibinfo {author} {\bibfnamefont {M.~H.}\ \bibnamefont
  {Mikkelsen}}, \bibinfo {author} {\bibfnamefont {N.~G.}\ \bibnamefont
  {Stoltz}}, \bibinfo {author} {\bibfnamefont {L.~A.}\ \bibnamefont {Coldren}},
  \ and\ \bibinfo {author} {\bibfnamefont {D.~D.}\ \bibnamefont {Awschalom}},\
  }\href {\doibase 10.1126/science.1154798} {\bibfield  {journal} {\bibinfo
  {journal} {Science}\ }\textbf {\bibinfo {volume} {320}},\ \bibinfo {pages}
  {349} (\bibinfo {year} {2008})}\BibitemShut {NoStop}%
\bibitem [{\citenamefont {Meixner}\ \emph {et~al.}(1992)\citenamefont
  {Meixner}, \citenamefont {Jefferson},\ and\ \citenamefont
  {Macfarlane}}]{PhysRevB.46.5912}%
  \BibitemOpen
  \bibfield  {author} {\bibinfo {author} {\bibfnamefont {A.}~\bibnamefont
  {Meixner}}, \bibinfo {author} {\bibfnamefont {C.}~\bibnamefont {Jefferson}},
  \ and\ \bibinfo {author} {\bibfnamefont {R.}~\bibnamefont {Macfarlane}},\
  }\href {\doibase 10.1103/PhysRevB.46.5912} {\bibfield  {journal} {\bibinfo
  {journal} {Phys. Rev. B}\ }\textbf {\bibinfo {volume} {46}},\ \bibinfo
  {pages} {5912} (\bibinfo {year} {1992})}\BibitemShut {NoStop}%
\bibitem [{\citenamefont {Liao}\ \emph {et~al.}(2014)\citenamefont {Liao},
  \citenamefont {Keitel},\ and\ \citenamefont
  {P\'alffy}}]{PhysRevLett.113.123602}%
  \BibitemOpen
  \bibfield  {author} {\bibinfo {author} {\bibfnamefont {W.-T.}\ \bibnamefont
  {Liao}}, \bibinfo {author} {\bibfnamefont {C.~H.}\ \bibnamefont {Keitel}}, \
  and\ \bibinfo {author} {\bibfnamefont {A.}~\bibnamefont {P\'alffy}},\ }\href
  {\doibase 10.1103/PhysRevLett.113.123602} {\bibfield  {journal} {\bibinfo
  {journal} {Phys. Rev. Lett.}\ }\textbf {\bibinfo {volume} {113}},\ \bibinfo
  {pages} {123602} (\bibinfo {year} {2014})}\BibitemShut {NoStop}%
\bibitem [{\citenamefont {Lauritzen}\ \emph {et~al.}(2011)\citenamefont
  {Lauritzen}, \citenamefont {Min\'a\ifmmode~\check{r}\else \v{r}\fi{}},
  \citenamefont {de~Riedmatten}, \citenamefont {Afzelius},\ and\ \citenamefont
  {Gisin}}]{Lauritzen2011}%
  \BibitemOpen
  \bibfield  {author} {\bibinfo {author} {\bibfnamefont {B.}~\bibnamefont
  {Lauritzen}}, \bibinfo {author} {\bibfnamefont {J.~c.~v.}\ \bibnamefont
  {Min\'a\ifmmode~\check{r}\else \v{r}\fi{}}}, \bibinfo {author} {\bibfnamefont
  {H.}~\bibnamefont {de~Riedmatten}}, \bibinfo {author} {\bibfnamefont
  {M.}~\bibnamefont {Afzelius}}, \ and\ \bibinfo {author} {\bibfnamefont
  {N.}~\bibnamefont {Gisin}},\ }\href {\doibase 10.1103/PhysRevA.83.012318}
  {\bibfield  {journal} {\bibinfo  {journal} {Phys. Rev. A}\ }\textbf {\bibinfo
  {volume} {83}},\ \bibinfo {pages} {012318} (\bibinfo {year}
  {2011})}\BibitemShut {NoStop}%
\bibitem [{\citenamefont {Dajczgewand}\ \emph {et~al.}(2014)\citenamefont
  {Dajczgewand}, \citenamefont {{Le Gou\"{e}t}}, \citenamefont
  {Louchet-Chauvet},\ and\ \citenamefont {Chaneli\`{e}re}}]{Dajczgewand14}%
  \BibitemOpen
  \bibfield  {author} {\bibinfo {author} {\bibfnamefont {J.}~\bibnamefont
  {Dajczgewand}}, \bibinfo {author} {\bibfnamefont {J.-L.}\ \bibnamefont {{Le
  Gou\"{e}t}}}, \bibinfo {author} {\bibfnamefont {A.}~\bibnamefont
  {Louchet-Chauvet}}, \ and\ \bibinfo {author} {\bibfnamefont {T.}~\bibnamefont
  {Chaneli\`{e}re}},\ }\href {\doibase 10.1364/OL.39.002711} {\bibfield
  {journal} {\bibinfo  {journal} {Opt. Lett.}\ }\textbf {\bibinfo {volume}
  {39}},\ \bibinfo {pages} {2711} (\bibinfo {year} {2014})}\BibitemShut
  {NoStop}%
\bibitem [{\citenamefont {Dajczgewand}\ \emph {et~al.}(2015)\citenamefont
  {Dajczgewand}, \citenamefont {Ahlefeldt}, \citenamefont {B\"ottger},
  \citenamefont {Louchet-Chauvet}, \citenamefont {Gou\"et},\ and\ \citenamefont
  {Chaneli\`ere}}]{dajczgewand2014optical}%
  \BibitemOpen
  \bibfield  {author} {\bibinfo {author} {\bibfnamefont {J.}~\bibnamefont
  {Dajczgewand}}, \bibinfo {author} {\bibfnamefont {R.}~\bibnamefont
  {Ahlefeldt}}, \bibinfo {author} {\bibfnamefont {T.}~\bibnamefont
  {B\"ottger}}, \bibinfo {author} {\bibfnamefont {A.}~\bibnamefont
  {Louchet-Chauvet}}, \bibinfo {author} {\bibfnamefont {J.-L.~L.}\ \bibnamefont
  {Gou\"et}}, \ and\ \bibinfo {author} {\bibfnamefont {T.}~\bibnamefont
  {Chaneli\`ere}},\ }\href {http://stacks.iop.org/1367-2630/17/i=2/a=023031}
  {\bibfield  {journal} {\bibinfo  {journal} {New J. Phys.}\ }\textbf {\bibinfo
  {volume} {17}},\ \bibinfo {pages} {023031} (\bibinfo {year}
  {2015})}\BibitemShut {NoStop}%
\bibitem [{\citenamefont {Allen}\ and\ \citenamefont
  {Eberly}(1987)}]{allen1987ora}%
  \BibitemOpen
  \bibfield  {author} {\bibinfo {author} {\bibfnamefont {L.}~\bibnamefont
  {Allen}}\ and\ \bibinfo {author} {\bibfnamefont {J.}~\bibnamefont {Eberly}},\
  }\href@noop {} {\emph {\bibinfo {title} {{Optical resonance and two-level
  atoms}}}}\ (\bibinfo  {publisher} {Courier Dover Publications},\ \bibinfo
  {year} {1987})\BibitemShut {NoStop}%
\bibitem [{\citenamefont {B\"ottger}\ \emph {et~al.}(2009)\citenamefont
  {B\"ottger}, \citenamefont {Thiel}, \citenamefont {Cone},\ and\ \citenamefont
  {Sun}}]{Bottger09}%
  \BibitemOpen
  \bibfield  {author} {\bibinfo {author} {\bibfnamefont {T.}~\bibnamefont
  {B\"ottger}}, \bibinfo {author} {\bibfnamefont {C.~W.}\ \bibnamefont
  {Thiel}}, \bibinfo {author} {\bibfnamefont {R.~L.}\ \bibnamefont {Cone}}, \
  and\ \bibinfo {author} {\bibfnamefont {Y.}~\bibnamefont {Sun}},\ }\href
  {\doibase 10.1103/PhysRevB.79.115104} {\bibfield  {journal} {\bibinfo
  {journal} {Phys. Rev. B}\ }\textbf {\bibinfo {volume} {79}},\ \bibinfo
  {pages} {115104} (\bibinfo {year} {2009})}\BibitemShut {NoStop}%
\end{thebibliography}%
%
%

%
\end{document}